\documentclass[final,3p,times,twocolumn]{elsarticle}  
\usepackage{amssymb}
\usepackage{color}
\usepackage{psfrag}
\usepackage{amsmath}
\usepackage{amssymb}
\usepackage{mathrsfs}
\usepackage[T1]{fontenc}
\usepackage{amsthm}

\DeclareMathAlphabet\mathbfcal{OMS}{cmsy}{b}{n}

\usepackage{units}
\theoremstyle{remark}

\usepackage[hyphens]{url}
\usepackage{multirow}
\usepackage{array}
\newcolumntype{L}[1]{>{\raggedright\let\newline\\\arraybackslash\hspace{0pt}}m{#1}}
\newcolumntype{C}[1]{>{\centering\let\newline\\\arraybackslash\hspace{0pt}}m{#1}}
\newcolumntype{R}[1]{>{\raggedleft\let\newline\\\arraybackslash\hspace{0pt}}m{#1}}

\usepackage{arydshln}

\usepackage{framed} 
\usepackage{multicol} 
\usepackage{nomencl} 
\makenomenclature
\renewcommand*\nompreamble{\begin{multicols}{2}}
\renewcommand*\nompostamble{\end{multicols}}
\usepackage{etoolbox}
\renewcommand\nomgroup[1]{%
  \item[\bfseries
  \ifstrequal{#1}{A}{Variables}{%
  \ifstrequal{#1}{B}{Indizes}{%
  \ifstrequal{#1}{C}{Superscripts}{}}}%
]}

\usepackage{tabularx}

\usepackage{pgfplots}
\pgfplotsset{compat=newest}
 
\usepackage{import} 
\usepackage{subcaption}

\usepackage{tikz}

\begin{document}

\begin{frontmatter}

\title{Supervisory model predictive control for PV battery and heat pump system with phase change slurry thermal storage}


\author[1]{Yannik L\"ohr} 
	\ead{yannik.loehr@rub.de}
\author[2]{Daniel Wolf} 
	\ead{daniel.wolf@homepowersolutions.de}
\author[3]{Clemens Pollerberg} 
   \ead{clemens.pollerberg@umsicht.fraunhofer.de}
\author[4]{Alexander H\"orsting} 
	\ead{alexander.hoersting@stiebel-eltron.de}
\author[1]{Martin M\"onnigmann\corref{cor1}}
	\ead{martin.moennigmann@rub.de}

\cortext[cor1]{Corresponding author}

\address[1]{Automatic Control and Systems Theory, Ruhr-Universit\"at Bochum, 44801 Bochum, Germany}
\address[2]{HPS Home Power Solutions GmbH, 12489 Berlin, Germany}
\address[3]{Fraunhofer UMSICHT, 46047 Oberhausen, Germany}
\address[4]{STIEBEL ELTRON GmbH \&  Co. KG, 37603 Holzminden, Germany}

\begin{abstract}
We present the design, implementation and experimental validation of a supervisory predictive control approach for an electrical heating system featuring a phase change slurry as heat storage and transfer medium. 
The controller optimizes the energy flows that are used as set points for the heat generation and energy distribution components. The optimization handles the thermal and electrical subsystems simultaneously and is able to switch between different objectives.
We show the control can be implemented on low-cost embedded hardware and validate it with an experimental test bed comprising an installation of the complete heating system, including all hydraulic and all electrical components.
Experimental results demonstrate the feasibility of both, a heat pump heating system with a phase change slurry, and the optimal control approach. 
The main control objectives, i.e., thermal comfort and maximum self-consumption of solar energy, can be met.
In addition, the system and its controller provide a load shifting potential.
\end{abstract}


\end{frontmatter}


\section{Introduction}
\label{sec:Intr}
\begin{table*}[t]
	\begin{framed}	
	\nomenclature[A]{$P$}{electrical power flow $(\unit{kW})$}
	\nomenclature[A]{$q$}{thermal power flow $(\unit{kW})$}
	\nomenclature[A]{$E$}{stored energy $(\unit{kWh})$}	
	\nomenclature[A]{$COP$}{heat pump coefficient of performance}
	\nomenclature[A]{$\alpha_{(\cdot)}$}{thermal loss coefficients}
	\nomenclature[A]{$\beta_{(\cdot)}$}{thermal efficiency coefficients}
	\nomenclature[A]{$G$}{global irradiation $(\unit{kW h / m^2})$}	
	\nomenclature[A]{$Y$}{installed peak power $(\unit{kW})$}	
	\nomenclature[A]{$r_{\star}$}{weight coefficients}
	\nomenclature[A]{$m$}{mass $(\unit{kg})$}
	\nomenclature[A]{$N_p$}{prediction horizon}
	\nomenclature[A]{$N_u$}{control horizon}	
	\nomenclature[A]{$c_p$}{specific heat capacity $(\unit{kJ / kg K})$}	
	\nomenclature[A]{$\rho$}{density $(\unit{kg / m^3})$}	
	\nomenclature[A]{$\dot{V}$}{volume flow $(\unit{l / s})$}
	\nomenclature[A]{$h$}{sampling time $(\unit{min})$}		
	\nomenclature[A]{$\eta$}{AC to DC power conversion coefficient}
	\nomenclature[A]{$\mu$}{PV power factor}
	\nomenclature[A]{$\mathbf{x}$}{states}
	\nomenclature[A]{$\mathbf{u}$}{inputs}
	\nomenclature[A]{$\mathbf{d}$}{disturbances}
	\nomenclature[A]{$\Delta z$}{height difference $(\unit{m})$}
	\nomenclature[A]{$z$}{height $(\unit{m})$}
	\nomenclature[A]{$p$}{pressure $(\unit{Pa})$}
	\nomenclature[A]{$\Delta h_f$}{specific enthalpy $(\unit{J / kg})$}	
	\nomenclature[A]{$\rm SoC$}{Battery state of charge $(\unit{\%})$}
	\nomenclature[A]{$w_\mathrm{P}$}{mass fraction $(\unit{\%})$}
	
	\nomenclature[B]{$\rm PV$}{photovoltaic}
	\nomenclature[B]{$\rm HP$}{heat pump}
	\nomenclature[B]{$\rm HR$}{heating rod}
	\nomenclature[B]{$\rm L$}{(household) load}
	\nomenclature[B]{$\rm DHW$}{domestic hot water}
	\nomenclature[B]{$\rm SH$}{space heating}
	\nomenclature[B]{$\rm B$}{battery}
	\nomenclature[B]{$\rm G$}{grid}
	\nomenclature[B]{$\rm W$}{water}
	\nomenclature[B]{$\rm PCS$}{phase change slurry}
	\nomenclature[B]{$\rm BLD$}{building mass}
	\nomenclature[B]{$\rm P$}{paraffin wax}
	\nomenclature[B]{$\rm sen$}{sensible}
	\nomenclature[B]{$\rm lat$}{latent}

	\nomenclature[C]{dem}{demand}
	\nomenclature[C]{sup}{supply}
	\nomenclature[C]{ch}{charging}
	\nomenclature[C]{dis}{discharging}
	\nomenclature[C]{ret}{return}

	\printnomenclature
	\end{framed}

	\vspace{-12pt}
\end{table*}
Buildings account for almost $40 \%$ of the global energy consumption \cite{kolokotsa2011}. 
Among numerous control concepts for buildings, model predictive control (MPC) has emerged as a promising alternative \cite{qin2003}. MPC is an optimal control strategy with two main features that are significant for building control, namely the minimization of a performance criterion and the systematic handling of constraints. The concept rests upon model-based predictions of the future system behavior. Combining a performance criterion, constraints and system model, an optimal control problem (OCP) is posed and solved to determine the optimal system behavior. The first element of the solution, i.e., the optimal control input, is then applied to the system, before the OCP is updated and solved again at the next time step \cite{clarke1991, mayne2000,afram2014}. 

MPC has been applied to heating, ventilation and air conditioning (HVAC) systems in buildings before. An overview is given in \cite{thieblemont2017}, where special attention was paid to buildings with energy storage systems and weather forecasts. 
Existing studies differ with respect to the main control objective.
In \cite{sevilla2015} for example, 
the self-consumption of energy generated by on-site photovoltaic system was maximized.  
Other authors incorporate economic aspects \cite{kuboth2019,oldewurtel2010} or treat demand side management \cite{arteconi2013, bianchini2016}.
Simulations of distributed economic MPC for an electrical heating system featuring thermal energy storage and electrical energy storage were presented in \cite{kuboth2019}.
In \cite{oldewurtel2010}, the authors used a real-time pricing scheme to achieve peak-shaving with HVAC systems. Their approach is based on flexible electricity prices, which, after being subject in scientific publications for two decades \cite{bjorgan2000,schreiber2015}, find the way into real end-consumer contracts \cite{reuters2017} today. 
In \cite{fiorentini2017}, the authors presented a hybrid model predictive controller for HVAC of a net zero energy building with phase change material (PCM) included in the air flow line of a heat pump. The experimental study was one of the first that combined predictive control with PCM in a building application. However, in contrast to the study reported in the present paper, standard PC hardware was used.

Phase change materials are an interesting alternative to water as a heat energy carrier in buildings \cite{shao2015}. PCM are characterized by a high specific thermal capacity around the temperature range of the phase change, because latent heat is stored in addition to sensible heat \cite{zalba2003}.
Current research on PCM ranges from applications of the phase change to effectively increasing thermal insulation of buildings to increasing the thermal capacity of heat storages. An overview of building related applications is given in \cite{kalnaes2015}. 
Phase change slurries (PCS) are an important class of PCM for HVAC systems \cite{zhang2010}. This class stands out, as these materials are both, heat storage and heat transfer medium. Technically, PCS are a mixture of water and a phase change material, where microscopically small particles of an arbitrary phase change material are, for example, dispersed or encapsulated in water \cite{delgado2012}.
PCS have first been used in buildings as ice slurries in cooling applications, see \cite{huang2009, huang2010}. In \cite{kappels2014}, the authors replaced water with PCS in a cold storage application. As a result, the cold storage capacity was increased by almost a factor of $3$, while volume flow and heat transfer remain sufficient without significant changes in pumping power.
\begin{figure*}[ht]	
	\center 
	\def\svgwidth{0.77\textwidth} 
	\small{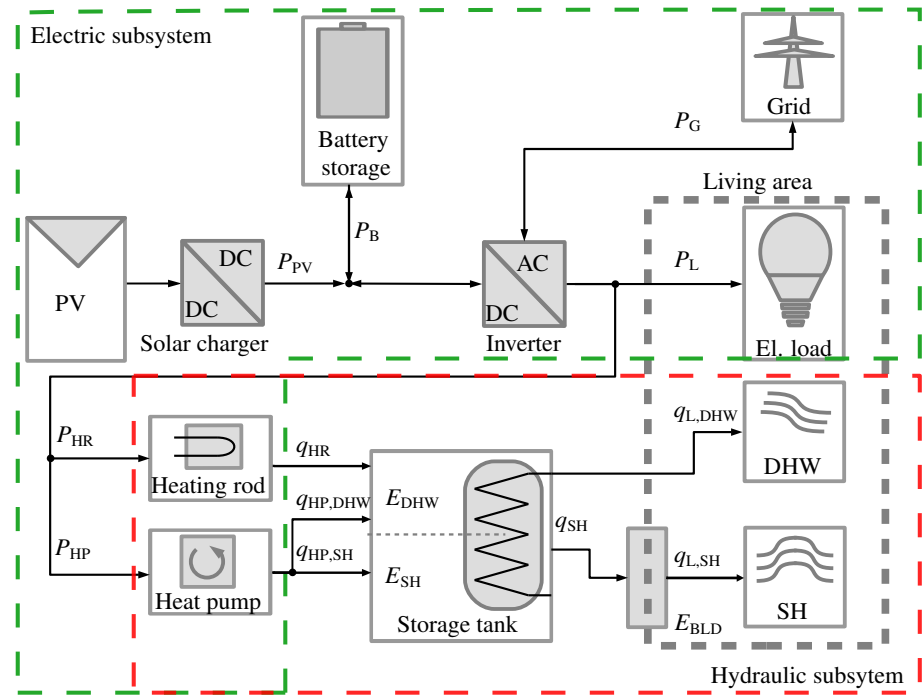}
	\vspace{-6pt}
	\caption{Sketch of the heat generation and consumption system. $P$ and $q$ denote power and heat flows, respectively. }
	\vspace{-6pt}
	\label{fig:SysStruct}
\end{figure*}

We introduce a novel phase change slurry with an adequate phase change temperature for low temperature heating systems. 
Since the PCS has a higher heat capacity than water, it results in a higher system heat capacity. 
Increasing the heat capacity is particularly desirable in domestic heating systems with renewable energy generation. It increases the potential to shift loads and enables a higher self-consumption fraction of solar energy. 
In order to use the system to full capacity and handle different operation objectives, a supervisory model predictive control approach for the electrical heating system is developed.
We show it can be implemented on low-cost embedded hardware that is suitable for domestic heating systems. 

Existing publications are either restricted to simulation studies or laboratory computational hardware. 
In contrast, it is one of our main contributions to prove an implementation on low-cost embedded hardware is possible.
The proposed predictive controller can be added to existing systems as a supervisory, or piggyback, controller.
The phase change slurry is used with standard hydraulic components, i.e., the only change is to replace water with the PCS.
The performance of the supervisory MPC is illustrated with experimental results. 

A control-oriented model for the 
electrical heating system is presented in Section \ref{sec:SysConfig}.
In Section \ref{sec:PredCtrl}, the supervisory MPC is proposed. The experimental test-bed is introduced in Section \ref{sec:RT_EmbdHW}, together with a brief explanation of the real-time implementation.
Experimental results are presented in Section \ref{sec:ExpRes}.

\section{System description and modeling}
\label{sec:SysConfig}   
The hydraulic and electric part of the system are introduced in Sections \ref{subsec:Hydr} and \ref{subsec:Elec}, respectively. They are combined into a control-oriented system model in Section \ref{subsec:ContrOrSysMod}, which uses adjustable heat and power flows as inputs.

\subsection{Introduction}
\label{subsec:Introduction}
The system treated here is sketched in Figure \ref{fig:SysStruct}.
Heat is generated by a heat pump and an auxiliary heating rod and stored in a heat storage tank that is filled with PCS.
An internal heat exchanger is used to provide domestic hot water (DHW), while space heating (SH) demand is met by discharging heat from the storage directly. 
The thermal mass of the building is treated as an additional thermal storage.

Electricity is provided by photovoltaic panels and a bidirectional connection to the electrical grid. 
Electricity required for heat generation and the baseline household consumption constitute the electric load demand.
The system includes an electrochemical battery.
The system has been partially introduced in \cite{loehr2017} and \cite{loehr2018b}. 

\subsection{Hydraulic subsystem}
\label{subsec:Hydr}
The hydraulic subsystem consists of a storage tank filled with a phase change slurry, a modulating air source heat pump and a heating rod (see Figure \ref{fig:SysStruct}). 
Heat is removed from the system with two heat exchangers, where, in both cases, the discharge is operated with water. 

\subsubsection{Latent heat storage with a phase change slurry}
\label{subsec:PCS}
The phase change slurry used here is an emulsion of paraffin wax particles dispersed in water \cite{youssef2013}. The paraffin wax is selected to have a phase change range adequate for floor heating.

\begin{figure}[ht]
\centering
\small
\def\svgwidth{0.47\textwidth} 
\small{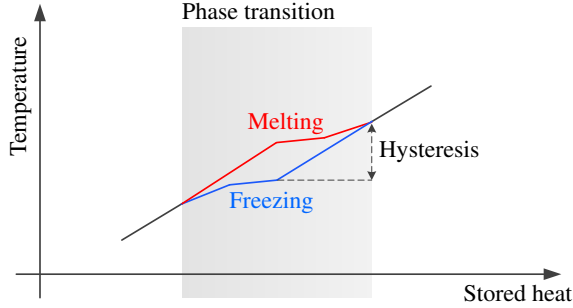}
\caption{Illustration of hysteresis in PCS phase transitions.}
\label{fig:hysteresis}
\end{figure}

The hysteresis sketched in Figure~\ref{fig:hysteresis} results, because the phase change of paraffin occurs in a different temperature range for freezing than for melting. 
During the phase transition, the gradient of the temperature change decreases and temperature plateaus result. 
Because the curve is nonunique in this range, the state of charge cannot be determined from a temperature measurement. 
We explain how the state of charge can be determined in the remainder of the section. 

The heat stored by the PCS in the storage tank amounts to
\begin{align}\nonumber
E_\mathrm{PCS} &= E_\mathrm{sen}+E_\mathrm{lat} \\ \label{eq:q_PCS}
&= m_\mathrm{PCS} \cdot \left[ c_{p,\mathrm{PCS}} \cdot \Delta T + w_\mathrm{P} \cdot \Delta h_f \right],
\end{align}
with specific heat capacity $c_{p,\mathrm{PCS}} =\left(1-w_\mathrm{P} \right) \cdot c_{p,\mathrm{w}}+w_\mathrm{P}\cdot c_{p,\mathrm{P}}$, where $w_\mathrm{P}$ is the mass fraction of paraffin wax in the PCS, $m_\mathrm{PCS}$ is the mass of PCS in the storage tank, $c_{p,\mathrm{w}}$ and $c_{p,\mathrm{P}}$ are the specific heat capacities of water and paraffin, respectively, $\Delta T$ is the usable temperature difference and $\Delta h_f$ is the enthalpy difference in the phase change of the paraffin. 
Parameters $w_\mathrm{P}$, $c_{p,\mathrm{w}}$ and $c_{p,\mathrm{P}}$ are design parameters or known from material data sheets. 

The stored sensible heat, i.e., $E_\mathrm{sen} = m_\mathrm{PCS} \cdot c_{p,\mathrm{PCS}} \cdot \Delta T$, can be calculated from temperature measurements.
In contrast, determining the stored latent heat requires additional measurements.
We choose to use two pressure measurements, because pressure sensors are inexpensive. The pressure $p_i$ at height $z_i$ in the storage tank respects $\rho g z_i= p_i$. 
Consequently, the density $\rho$ can be calculated from  
\begin{equation} \label{eq:rho_from_deltaP}
\rho = \frac{\Delta p}{g \cdot \Delta z},
\end{equation}
where $\Delta p = p_\mathrm{bottom}-p_\mathrm{top}$ and $\Delta z= z_\mathrm{bottom}- z_\mathrm{top}$ refer to two measurements close to the bottom and top of the storage tank. 
Now let $\rho_\mathrm{PCS,liq}$ and $\rho_\mathrm{PCS,sol}$ refer to the density of the paraffin water emulsion if the paraffin is completely liquid and solid, respectively. The density $\rho$ of the emulsion then lies in the interval $\rho\in [\rho_\mathrm{PCS,liq}, \rho_\mathrm{PCS,sol}]$ and
\begin{align*}
  \frac{\rho- \rho_\mathrm{PCS,sol}}{\rho_\mathrm{PCS,liq}-\rho_\mathrm{PCS,sol}}\in [0, 1]
\end{align*}
indicates the fraction of liquid paraffin. 
Assuming the density difference is proportional to the latent heat $E_\mathrm{lat}$ in the tank,
the stored latent heat can be determined from
\begin{equation}\label{eq:latent_heat}
E_\mathrm{lat}\left( \rho\right) = m_\mathrm{PCS}\, w_\mathrm{P}\,\Delta h_f\frac{\rho- \rho_\mathrm{PCS,sol}}{\rho_\mathrm{PCS,liq}-\rho_\mathrm{PCS,sol}},
\end{equation}
where $\rho$ is determined from two pressure measurements according to~\eqref{eq:rho_from_deltaP}. 
The densities $\rho_\mathrm{PCS,liq}$ and $\rho_\mathrm{PCS,sol}$ can be obtained from simple measurements.

\subsubsection{Heat consumption and heat generation}
\label{subsec:HeatConHeatGen}
\begin{figure}[ht]
\centering
\def\svgwidth{0.475\textwidth}
\small{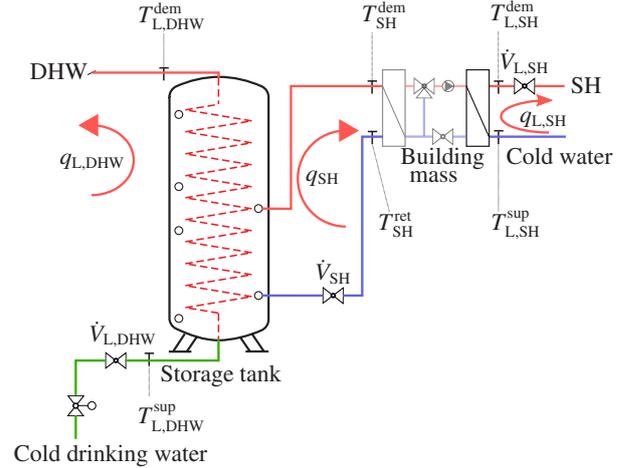}
\vspace{-10pt}
\caption{Heat consumption circuit.}
\vspace{-10pt}
\label{fig:th_load_schematic}
\end{figure}

The heat consumption subsystem is sketched in Figure \ref{fig:th_load_schematic}, where $q_{\rm L,SH}$ and $q_{\rm L,DHW}$ represent the SH and DHW load, respectively, introduced in Figure \ref{fig:SysStruct}.
While the DHW demand is satisfied using an internal heat exchanger within the storage tank, space heating is delivered via an external heat exchanger decoupling heat generation from heat distribution. The intermediate circuit shown grayed out is required for physical simulation of a floor heating.
The heat consumption can be calculated from
\begin{equation}
\label{eq:q_L_SH}
q_{\rm L,\star} = \rho_{\rm w} \cdot \dot{V}_{\rm L,\star } \cdot c_{\rm p,w} \cdot \left( T_{\rm L,\star }^{\rm dem} - T_{\rm L,\star }^{\rm sup} \right), \star \in \left\lbrace \text{SH,DHW} \right\rbrace.
\end{equation}

The heat pump is operated at a SH and DHW mode. The modes specify if the heat pump generates heat on appropriate temperature levels for SH or DHW supply, respectively. The corresponding heat flows are $q_\mathrm{HP,SH}$ and $q_\mathrm{HP,DHW}$.
The second heat generator is an auxiliary heating rod located in the upper part of the storage that provides the heat flow $q_\mathrm{HR}$.
We anticipate that these values serve as inputs in the dynamic model presented in Section \ref{subsec:ContrOrSysMod} and omit a detailed description here. Additional information is given in \ref{sec:app:heat_gen_and_release_PCS}.

\subsection{Electric subsystem}
\label{subsec:Elec}
\begin{figure}[ht]
\centering
\def\svgwidth{0.47\textwidth}
\small{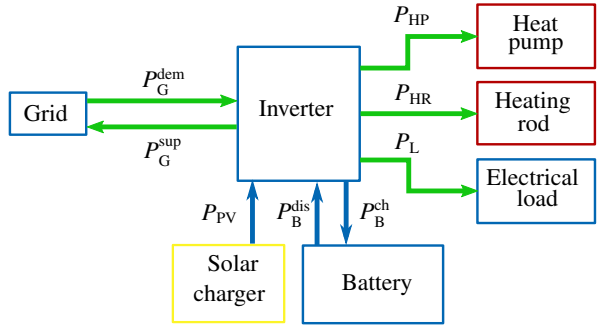}
\vspace{-5pt}
\caption{Sketch of electrical network. Green and blue lines depict AC and DC connections, respectively.}
\label{fig:electrical_network}
\end{figure}

The electric subsystem consists of a photovoltaic power generation, a grid connection and a battery storage. It supplies the demand required for heat generation and the household electric load. Figure \ref{fig:electrical_network} shows the electrical connections in the system, where the interaction with the grid $P_\mathrm{G}$ and the battery $P_\mathrm{B}$ are both bidirectional. The maximum capacity of the battery is $E_\mathrm{B}^\mathrm{max}$ and the current amount of electrical energy stored is 
\begin{equation} \label{eq:BatterySoC}
E_\mathrm{B} = \text{SoC} \cdot E_\mathrm{B}^\mathrm{max}.
\end{equation}
The electrical energy demand required for heat generation reads
\begin{eqnarray}\label{eq:ThermElecCoupling} \label{eq:P_HP}
P_{\rm HP} &=& \frac{q_{\rm HP,SH}}{COP_{\rm SH}} + \frac{q_\mathrm{HP,DHW}}{COP_{\rm DHW}}, \\
P_{\rm HR} &=& q_{\rm HR}.
\end{eqnarray}
The photovoltaic power output is calculated from global irradiation $G$ using the simple relationship for a surface perpendicular to direct solar radiation,
\begin{equation}\label{eq:P_PV} 
P_{\rm PV} = \frac{G}{1000 \frac{W}{m^2}} \cdot Y_{\rm PV} \cdot \mu,
\end{equation}
where $Y_{\rm PV}$ the installed photovoltaic peak power at standard test conditions (STC) and $\mu \in \left( 0, 1 \right)$ a soiling factor. 
The household load $P_\mathrm{L}$ is assumed to be measurable throughout the paper.
According to Kirchhoff's point rule, the sum of power into the electrical node labeled \textit{Inverter} in Figure \ref{fig:electrical_network} must always be equal to zero, i.e., 
\begin{equation}\label{eq:ElecCoupling}
P_{\rm G}^{\rm dem}+\eta P_{\rm B}^{\rm dis}+\eta P_{\rm PV} = P_\mathrm{L}+\eta P_{\rm B}^{\rm ch}+P_{\rm G}^{\rm sup}+P_{\rm HP}+P_{\rm HR},
\end{equation}
where $\eta$ collects the conversion efficiency. 

\subsection{Control-oriented system model}
\label{subsec:ContrOrSysMod}
Heat generation and heat withdrawal from the storage are nonlinear relationships, if the temperatures and volume flows are used as states and inputs.
In contrast, we propose to use a linear discrete-time representation of the change in the stored energy and to collect physical properties like losses and efficiencies by identifiable parameters.
In the following, $k=0,1,2,\dots$ enumerates discrete time steps.
The stored energies are considered as state variables, i.e., $\mathbf{x}(k) = \left[E_{\rm SH}(k),E_{\rm DHW}(k),E_{\rm BLD}(k), E_{\rm B}(k)\right]^T$.

The thermal and electrical energy flows constitute the inputs
\begin{equation} \label{eq:EnergyFlows}
\mathbf{u}(k) = \begin{bmatrix}
q_{\rm HP,SH}(k)\\
q_\mathrm{HP,DHW}(k)\\
 q_{\rm HR}(k)\\
 q_{\rm SH}(k)\\
 P_{\rm B}^{\rm ch}(k)\\
 P_{\rm B}^{\rm dis}(k)\\
 P_{\rm N}^{\rm dem}(k)\\
 P_{\rm N}^{\rm sup}(k)
\end{bmatrix}.
\end{equation}
The thermal load demand are considered to be measured disturbances $\mathbf{d}(k) = \left[q_{\rm L,SH}(k), \allowbreak q_{\rm L,DHW}(k)\right]$ and modeled according to~\eqref{eq:q_L_SH}.

The two parts of the thermal storage are modeled with thermally connected, simplified energy balance equations.
With $E_{\rm SH}(k)$ and $E_{\rm DHW}(k)$ as absolute value of stored thermal energy in lower and upper part of the storage, the linear discrete-time dynamic equations read
 \begin{subequations}\label{eq:HeatStorageDynamics}
 \begin{alignat}{2}\label{eq:HeatStorageDynamics_low}
 \small
 E_{\rm SH}(k+1) &= \alpha_1 E_{\rm SH}(k) +\nu E_{\rm DHW}(k)\\ \nonumber
 &+\beta_1 q_{\rm HP,SH}(k) - \beta_2 q_{\rm SH}(k), \\ 
 \label{eq:HeatStorageDynamics_up}
 \small
  E_{\rm DHW}(k+1) &=  \left(\alpha_2-\nu\right) E_{\rm DHW}(k) \\ \nonumber
  &+\beta_3 q_{\rm HP, DHW}(k) +\beta_4 q_{\rm HR}(k) - \beta_5 q_{\rm L,DHW},
\end{alignat} 
 \end{subequations}
where the dimensionless parameters $\alpha_i,i=1,2$, $\nu$ and $\beta_i, i = 1,\dots,5$ describe losses and efficiencies, respectively.
The dynamics of the energy stored in the thermal building mass are described by 
\begin{equation}\label{eq:therm_load_mismatch}
E_{\rm BLD}(k+1) = \alpha_3 E_{\rm BLD}(k)+\beta_6\left(q_{\rm SH}(k)-q_{\rm L,SH}(k)\right),
\end{equation}
which can be understood as the dynamics of the difference between thermal load demand and heat drawn from the storage. 
The simplified energy balance of the electrical energy stored in the battery reads
\begin{equation}\label{eq:BatteryDynamics}
E_{\rm B}(k+1) = \alpha_4 E_{\rm B}(k) + \beta_7 P_{\rm B}^{\rm ch}(k)-\beta_8 P_{\rm B}^{\rm dis}(k),
\end{equation}
where $0 < \beta_7 < 1$ and $\beta_8 > 1$ are charging and discharging efficiencies, respectively.
The parameters in~\eqref{eq:HeatStorageDynamics}-\eqref{eq:BatteryDynamics} are derived from measurements and grey box estimation techniques (see \ref{sec:app:parameter_identification}).

Collecting~\eqref{eq:HeatStorageDynamics},~\eqref{eq:therm_load_mismatch} and~\eqref{eq:BatteryDynamics} yields the linear discrete-time state space model 
\begin{eqnarray}\label{eq:OvSysModel}
\nonumber
\mathbf{x}(k+1) &=& \mathbf{A}\mathbf{x}(k)+\mathbf{B}\mathbf{u}(k)+\mathbf{E}\mathbf{d}(k) \\
\mathbf{y}(k) &=& \mathbf{C} \mathbf{x}(k),
\end{eqnarray}
with system matrices
\[
  	\mathbf{A} = \begin{bmatrix}
	\alpha_1 & \nu & 0 & 0  \\
	0 & (\alpha_2-\nu) & 0 & 0 \\
	0 & 0 & \alpha_3 & 0 \\
	0 & 0 & 0 & \alpha_4
  	\end{bmatrix},
\]
\[
	\mathbf{B} = \begin{bmatrix}
	\beta_{1} & 0 & -\beta_{2} & 0 & 0 & 0 & 0 & 0 \\
	0 & \beta_{3}  & \beta_{4} & 0 & 0 & 0 & 0 & 0 \\
	0 & 0 & 0 & \beta_{6} & 0 & 0 & 0 & 0 \\
	0 & 0 & 0 & 0 & \beta_7 & -\beta_8 & 0 & 0 \\
   	\end{bmatrix}
   	\mbox{,}
\]
\[
  	\mathbf{E} = \begin{bmatrix}
  	0 & 0 \\
  	0 & -\beta_{5} \\
  	-\beta_{6} & 0\\
  	0 & 0
  	\end{bmatrix}
  	\mbox { and } 
  	\mathbf{C} = \left[ \mathbf{I}_{4 \times 4} \right].
\]
The states are calculated from~\eqref{eq:E_SH},~\eqref{eq:E_DHW} and~\eqref{eq:BatterySoC}. 

\section{Supervisory model predictive control}
\label{sec:PredCtrl}
%

It is the purpose of the control to determine optimal values for the energy flows~\eqref{eq:EnergyFlows}. The corresponding optimization problem is introduced in Section \ref{subsec:mpc_design}. 
We stress again the merit function to be optimized can be selected by the user to switch between various objectives, such as maximum PV self-consumption or minimum short-term operation cost. 
The MPC is a supervisory controller in the sense that the optimal energy flows~\eqref{eq:EnergyFlows} need to be implemented by lower level controllers. An example for such a lower level controller is a pump controller that adjusts $\dot{V}_\mathrm{HP}$ to achieve the required $q_\mathrm{HP}$.

Section \ref{subsec:implementational_aspects} discusses implementational aspects for  real-time and safe system operation. In Section \ref{subsec:weath_load_forec}, we give further information about the methodology used for forecasting $\mathbf{d}(k)$.  

\subsection{MPC problem}
\label{subsec:mpc_design}

MPC requires to solve the following optimal control problem:
\begin{subequations}
	\label{eq:OCP}
	\begin{align}
		\min _{\mathbf{U}(k)} \; & \; J(k) = J_{y}(k) + J_{u}(k) \label{eq:OCP:obj}\\
		\text{s.t.} \; & \; \mathbf{x}(k+1) = \mathbf{A} \mathbf{x}(k) + \mathbf{B} \mathbf{u}(k) + \mathbf{E} \mathbf{d}(k), \label{eq:OCP:x}\\
                      & \; E_{ j_1}^{\rm min} \leq E_{j_1 }(k) \leq E_{j_1 }^{\rm max}, \label{eq:OCP:StorageBounds}\\
		               & \; 0 \leq q_{j_2}(k) \leq q_{j_2}^{\rm max}(k), \label{eq:OCP:con_heatgen} \\
		               & \; |q_{\rm SH}(k)-q_{\rm L,SH}(k)| \leq \sigma_{th}(k), \label{eq:OCP:con_heatcon}  \\
		               & \; P_{j_3}^{j_4,\rm min}(k) \leq P_{j_3}^{j_4}(k) \leq P_{j_3}^{j_4,\rm max}(k), \label{eq:OCP:con_elecpower} \\
		               &~\eqref{eq:ElecCoupling}, \label{eq:OCP:eleccoupling}
	\end{align}
\end{subequations}
where $j_1 \in \left\lbrace \mathrm{SH, DHW, BLD, B} \right\rbrace$, $j_2 \in \left\lbrace \mathrm{HP, HR, SH} \right\rbrace$, $j_3 \in \left\lbrace \mathrm{ B, G} \right\rbrace$ and $j_4 \in \left\lbrace \mathrm{ch,dis,dem,sup} \right\rbrace$ refer to the corresponding system components and where $N$ is the number of simulated future time steps (the prediction horizon).
Solving~\eqref{eq:OCP} results in the optimal sequence $\mathbf{U}(k) = \left[\mathbf{u}(k|k), \mathbf{u}(k+1|k), \dots, \mathbf{u}(k+N|k) \right]$ of the input variables for the $N$ subsequent time steps. 
We anticipate that~\eqref{eq:OCP} is solved with a sampling time $h = 15~\unit{min}$. Consequently, every 15 minutes, the optimal values of~\eqref{eq:EnergyFlows} are updated.
The notation $\mathbf{u}(k+i|k)$ etc. refers to the predicted values of $\mathbf{u}(k)$ for time $k+i$ based on the information available at time $k$. 

The two terms of cost function~\eqref{eq:OCP:obj} are
{\small
\begin{align}\label{eq:CostFunction_Outputs}
\scriptsize
\nonumber
J_{y}(k) = \sum_{i=0}^{N}& r_{\rm E,SH}(k+i)\left(E_{\rm SH}^{\rm set}(k+i) - E_{\rm SH}(k+i|k)\right)^2 \\
\nonumber
& + r_{\rm E,DHW}(k+i)\left(E_{\rm DHW}^{\rm set}(k+i) - E_{\rm DHW}(k+i|k)\right)^2 \\
\nonumber
& + r_\mathrm{E,BLD}(k+i)\left(E_{\rm BLD}(k+i|k)\right)^2 \\
& + r_{\rm E,B}(k+i)\left(E_{\rm B}^{\rm set}(k+i) - E_{\rm B}(k+i|k)\right)^2
\end{align}
}
and
{\small
\begin{align}\label{eq:CostFunction_Inputs}
\scriptsize
\nonumber
J_{u}(k) = \sum_{i=0}^{N}& r_{\rm HP,SH}(k+i)\left(q_{\rm HP,SH}(k+i|k)\right)^2 \\
\nonumber
& + r_\mathrm{HP,DHW}(k+i)\left(q_\mathrm{HP,DHW}(k+i|k)\right)^2 \\
\nonumber
& + r_{\rm HR}(k+i)\left(q_{\rm HR}(k+i|k)\right)^2 \\
\nonumber
& + r_{\rm B,ch}(k+i)\left(P_{\rm B}^\mathrm{ch}(k+i|k)\right)^2 \\
\nonumber
& + r_{\rm B,dis}(k+i)\left(P_{\rm B}^\mathrm{dis}(k+i|k)\right)^2 \\
\nonumber
& + r_{\rm G,dem}(k+i)\left(P_{\rm G}^\mathrm{dem}(k+i|k)\right)^2\\
& + r_{\rm G,sup}(k+i)\left(P_{\rm G}^\mathrm{sup}(k+i|k)\right)^2,
\end{align}
}
which penalize the control error and input values, respectively, with flexible tuning parameters $r_{(\star)}(k+i) \geq 0$. 
Large values of these parameters in~\eqref{eq:CostFunction_Outputs} enforce small control errors, i.e., small differences between the control output and their set points. Similarly, large values in~\eqref{eq:CostFunction_Inputs} enforce small values of the corresponding energy flows.
Hence, maximum PV-self consumption, i.e., isolating the system from the grid, is achieved by assigning a high penalty $r_{\rm G,sup}(k)$ on grid supply and low penalties $r_{\rm B,\left(ch,dis\right)}(k)$ on battery actions. This is supported by assigning $E_{\rm B}^{\rm set} = E_{\rm B}^{\rm max}$ in~\eqref{eq:CostFunction_Outputs} combined with a high penalty $r_{\rm E,B}(k)$ during the day to prefer high level of stored electrical energy.
Other control objectives like minimizing the short-term operation cost or the energy consumption can be achieved in a similar manner.
Specific tuning parameters will be given in Section \ref{sec:RT_EmbdHW}.

Constraints~\eqref{eq:OCP:StorageBounds}-\eqref{eq:OCP:eleccoupling} restrict states and inputs, respectively. 
For example,~\eqref{eq:OCP:StorageBounds} ensures the lower part of the thermal storage to be operated only within the phase change range. This can be achieved by setting $E_\mathrm{SH}^{\rm min} = 0$ and $E_\mathrm{SH}^{\rm mix}$ according to the energy that is stored if the tank temperature is equal to the upper end of the phase transition range, compare Figure \ref{fig:hysteresis}.  
Similarly,~\eqref{eq:OCP:StorageBounds} also ensures
\begin{itemize}
\item that the upper part of the thermal storage is always at a temperature above the minimum usable temperature for DHW supply,
\item that the over- and undersupply is limited
\item and that the battery state of charge stays above a threshold.
\end{itemize}
Constraint \eqref{eq:OCP:con_heatgen} enforces an upper bound on the generated heat, where $q_{\rm HP}^{\rm max}(k)$, depends on ambient conditions and on the coefficient of performance.
Condition~\eqref{eq:OCP:con_heatcon} ensures that SH load $q_\mathrm{L,SH}$ is fulfilled.
$P_\mathrm{B}^\mathrm{\left(ch,dis \right)}$ and $P_\mathrm{G}^\mathrm{\left(sup,dem \right)}$ must respect negative lower and positive upper bounds, as given in~\eqref{eq:OCP:con_elecpower}.
The last constraint~\eqref{eq:OCP:eleccoupling} implements node equation~\eqref{eq:ElecCoupling}. 
Specific values for all bounds will be stated in Section \ref{sec:RT_EmbdHW} again.

The OCP~\eqref{eq:OCP} can be transformed into a quadratic program of the form
\begin{eqnarray}\label{eq:QP}
\nonumber
 \min_{\mathbf{U}(k)} & \mathbf{U}^T(k) \mathbfcal{H}(k)\mathbf{U}(k)- \mathbfcal{F}^T(k)\mathbf{U}(k) \\
 \mbox{s.t.} & \mathbf{\Omega}(k) \mathbf{U} (k) \leq \boldsymbol{\omega}(k).
\end{eqnarray}
The first element of the solution, $\mathbf{u}(k|k) = [q_{\rm HP,SH}(k|k), \allowbreak q_\mathrm{HP,DHW}(k|k), \allowbreak q_{\rm HR}(k|k),\dots]^T$, i.e., the optimal energy flows for the current time step, are applied to the system.
As stated above,~\eqref{eq:QP} is updated periodically with sampling time $h$ with current values for disturbances and tuning parameters, and a new solution is obtained based on the prediction along horizon $N$. Only the first predicted optimal input signal is applied and subsequent signals are updated in the next sampling period.  
Details on the derivation of~\eqref{eq:QP} can be found, e.g., in \cite{maciejowski2002}.

\subsection{Implementational aspects}
\label{subsec:implementational_aspects}
We discuss some implementational aspects required to apply the presented supervisory MPC in this section. We omit the details and refer to literature on the methods whenever possible for brevity.

Including equality constraint~\eqref{eq:OCP:eleccoupling} is numerically challenging. In order to solve~\eqref{eq:QP} robustly, a slack variable $\epsilon_{el}=0.1~\unit{W}$ is introduced. 
Additionally, 1-norm constraint softening is performed on output constraints~\eqref{eq:OCP:StorageBounds} \cite[pp. 97-99]{maciejowski2002}. This method adds a slack variable to the constraint that allows for a certain constraint violation. The slack variable is penalized in the cost function, such that constraint violations only occur if necessary.
Furthermore, a move-blocking approach is implemented to reduce the number of optimization variables in later time-steps. This permits increasing the prediction horizon $N$~\cite{cagienard2004,barlow2010}.

Some properties of the heat pump and the heating rod are not captured by the linear modeling approach. The operation of the heating rod is limited to discrete stages. 
The heat pump must be operated above a minimum power threshold $q_\mathrm{HP}^\mathrm{min} > 0$ if switched on. The heat pump is also subject to minimum up and down times, i.e., it must be activated or deactivated for a minimum time.
Including these properties in the OCP would require integer decision variables \cite{ferraritrecate2004}, which currently renders the approach computationally intractable for embedded implementation.
The components automatically choose the closest realizable state to the thermal set-point. 
Hence, the electrical set-points are not adjusted correctly, which would lead to values for battery or grid power that differ from the calculated optimal ones.
As a remedy, a set of rules were developed as suitable post processing to adjust $P_\mathrm{B}$ and $P_\mathrm{G}$,  if one of the optimal values for $q_\mathrm{HP,SH}$, $q_\mathrm{HP,DHW}$ or $q_\mathrm{HR}$ cannot be realized.
Specifically, $q_\mathrm{HP}$ is set to zero if $0 < q_\mathrm{HP} < q_\mathrm{HP}^\mathrm{min}$ and $q_\mathrm{HR}$ is adjusted to the nearest possible stage below the optimal set-point. 
Consequently, there is an amount of electric power, $P_\mathrm{add}$, that was planned for heat generation but will not be consumed from the heat generators.
We add $P_\mathrm{add}$ to $P_\mathrm{B}^\mathrm{ch}$ if $\rm{SoC}_\mathrm{B} < 90 \%$ in case of $P_\mathrm{PV} \geq P_\mathrm{add}$. If $\rm{SoC}_\mathrm{B} \geq 90 \%$, $P_\mathrm{add}$ is supplied to the grid. If $P_\mathrm{PV} <  P_\mathrm{add}$, we first reduce $P_\mathrm{G}^\mathrm{dem}$ and then $P_\mathrm{B}^\mathrm{dis}$ accordingly.

Furthermore, we record current up and down times of the heat pump. If the controller demands for $q_\mathrm{HP} > q_\mathrm{HP}^\mathrm{min}$ but the minimum down-time is not reached, the power is redistributed as described above. In the opposite case, i.e., if the heat pump needs to run longer than desired, we increase $P_\mathrm{G}^\mathrm{dem}$ to fill the power gap required to continue running the heat pump. 

\subsection{Disturbance predictions}
\label{subsec:weath_load_forec}
The solution of MPC problem~\eqref{eq:OCP} requires to predict loads and PV power generation along the prediction horizon $N$. 
For sake of simplicity, we do not model the dynamics of these quantities. Instead, we derive them from available measurements (see Section \ref{subsec:HeatConHeatGen} and Section \ref{subsec:Elec}).

Consequently, predictions are obtained based on historical measurement data. To create the historical data, at each time step, $q_\mathrm{L,SH}$ and $q_\mathrm{L,DHW}$ are calculated by~\eqref{eq:q_L_SH} and $P_\mathrm{L}$ is measured directly. 
The values are stored in time series, which include data of the past seven days.
During run-time, the controller accesses the data stored for the upcoming hours of the same weekday of the past week and uses it as a forecast, which conforms with standard occupancy profiles \cite{oldewurtel2013b}.
A drawback of this method is that large differences in ambient temperatures or user behavior between two consecutive weeks reduce the quality of the predictions.
The forecast of the PV power output is calculated by~\eqref{eq:P_PV} based on global irradiation forecast data provided by Germany's National Meteorological Service (DWD) \cite{dwd2019}.


\section{Experimental test-bed and controller implementation}
\label{sec:RT_EmbdHW}

\subsection{Experimental setup of the test-bed}
\label{sec:ExpSetup}
\begin{figure}[ht]
\center
\def\svgwidth{0.47\textwidth} 
\scriptsize{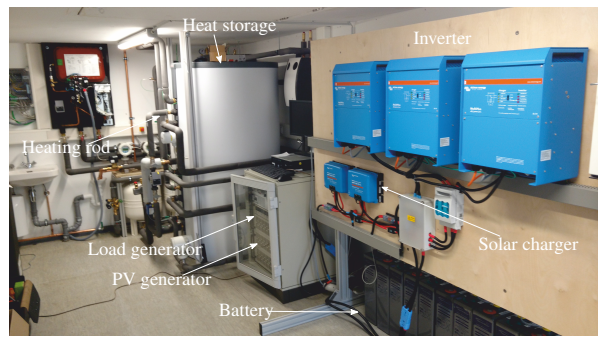}
\caption{Experimental test bed with heat storage, hydraulic piping, control station, inverter, solar charger and batteries. The heat pump is installed outside the building. 
The energy management hardware is installed on the back side of the wooden rack.}
\label{fig:exp_test_bed}
\vspace{-10pt}
\end{figure}

The experimental test-bed is shown in Figure \ref{fig:exp_test_bed}\footnote{The test-bed was installed in Holzminden, Germany.}. It consists of a complete installation of the hydraulic and electric components introduced in Figure \ref{fig:SysStruct} and specified in Table \ref{tab:plant_dimensions}. Detailed information on the components is given in Section \ref{subsec:StorageComponents}. Thermal and electrical load demand are physically simulated by withdrawal of heat and a controllable electric load, respectively.

Photovoltaic power generation of a PV installation with a peak power of $6~\unit{kW}$ is mimicked by a controllable AC/DC converter and solar charge controllers. The trend of the potential PV power is determined with~\eqref{eq:P_PV} using irradiation measurements obtained with a sensor that is installed on the roof of the building. 
This way, we are able to adjust the PV array tilt and orientation by software in order to simulate that the direct radiation is always orthogonal to the PV array. Using a point-wise measurement instead of a real PV array, the effect of potential panel shading could be avoided. 
The electrical components are operated by an embedded energy management platform, which will be introduced in Section \ref{subsec:RT_impl}. In addition, this hardware platform gathers all measurements and runs the supervisory MPC.


\begin{table*}[ht]
\center
\small
\caption{Dimension of energy storage, heat generation and power distribution components and corresponding constraints in \eqref{eq:OCP}.}
\begin{tabular}{|l|ll|ll|}
\hline
\textbf{Item} & \textbf{Operating range} & \textbf{Capacity} & \textbf{Lower bound} & \textbf{Upper bound}\\
\hline
Battery  & $[35, 100]\%~\unit{SoC}$ & $21~\unit{kWh}$ & $E_\mathrm{B}^\mathrm{min} = 7.35~\unit{kWh}$ & $E_\mathrm{B}^\mathrm{max} = 21~\unit{kWh}$ \\
SH storage & $[24, 42]~^\circ\unit{C}$ & $300~\unit{l}$ & $E_\mathrm{SH}^\mathrm{min} = 0~\unit{kWh}$ & $E_\mathrm{SH}^\mathrm{max} = 8.4~\unit{kWh}$  \\
DHW storage & $[50, 65]~^\circ\unit{C}$ & $300~\unit{l}$ & $E_\mathrm{DHW}^\mathrm{min} = 0~\unit{kWh}$ & $E_\mathrm{DHW}^\mathrm{max} = 3.6~\unit{kWh}$  \\
\hline
Heat pump & $\left\lbrace 0, \left[1, 3.7\right] \right\rbrace~\unit{kW_\mathrm{el}}$ & & $q_\mathrm{HP,\left( SH, DHW \right)}^\mathrm{min} = 0~\unit{kW_\mathrm{th}}$ & $q_\mathrm{HP,\left( SH, DHW \right)}^\mathrm{max} = 11.1~\unit{kW_\mathrm{th}}$  \\
Heating rod & $\left\lbrace 0, 2, 4, 6 \right\rbrace~\unit{kW_\mathrm{el}}$& & $q_\mathrm{HR}^\mathrm{min} = 0~\unit{kW_\mathrm{th}}$ & $q_\mathrm{HR}^\mathrm{max} = 6~\unit{kW_\mathrm{th}}$ \\
Battery charging & $\left[0, 7 \right]~\unit{kW_\mathrm{el}}$ & & $q_\mathrm{B}^\mathrm{ch,min} = 0~\unit{kW_\mathrm{el}}$ & $q_\mathrm{B}^\mathrm{ch,max} = 7~\unit{kW_\mathrm{el}}$  \\
Battery discharging & $\left[0, 7 \right]~\unit{kW_\mathrm{el}}$& & $q_\mathrm{B}^\mathrm{dis,min} = 0~\unit{kW_\mathrm{el}}$ & $q_\mathrm{B}^\mathrm{dis,max} = 7~\unit{kW_\mathrm{el}}$  \\
Grid demand & $\left[0, 7.5 \right]~\unit{kW_\mathrm{el}}$ & & $q_\mathrm{G}^\mathrm{dem,min} = 0~\unit{kW_\mathrm{el}}$ & $q_\mathrm{G}^\mathrm{dem,max} = 7.5~\unit{kW_\mathrm{el}}$  \\
Grid supply & $\left[0, 7.5 \right]~\unit{kW_\mathrm{el}}$ & & $q_\mathrm{G}^\mathrm{sup,min} = 0~\unit{kW_\mathrm{el}}$ & $q_\mathrm{G}^\mathrm{sup,max} = 7.5~\unit{kW_\mathrm{el}}$ \\
\hline
\end{tabular}
\label{tab:plant_dimensions}
\end{table*}

\subsection{Energy storage components} \label{subsec:StorageComponents}
The electric battery is of tubular gel lead acid technology. It is operated above $35\%$ state of charge (SoC) to avoid deep discharge potentially leading to a reduction of the usable capacity due to sulphation. 
The thermal storage capacity as stated in Table \ref{tab:plant_dimensions} results for the lowest temperatures usable for SH and DHW supply, which amount to $24~^\circ C$ and $50~^\circ C$, respectively. 
In the SH storage, the temperature range includes the whole phase change in both directions, explaining the higher heat capacity in spite of a smaller temperature gradient compared to the DHW storage. 

\begin{figure}[ht]
	\small
	\center
	\def\svgwidth{0.45\textwidth} 
	\scriptsize{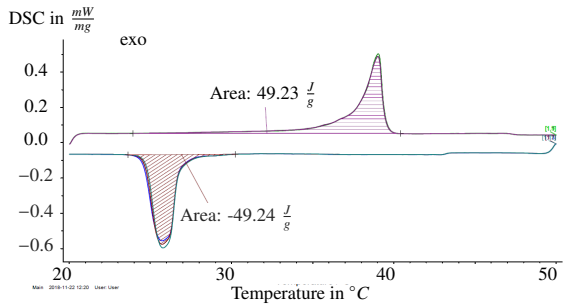}
	\captionof{figure}{Results of a DSC measurement for the used PCS.}
	\vspace{-6pt}
	\label{fig:dsc_meas}
\end{figure}

\begin{figure}[ht]
	\center	
	\small
	\def\svgwidth{0.3\textwidth} 
	\scriptsize{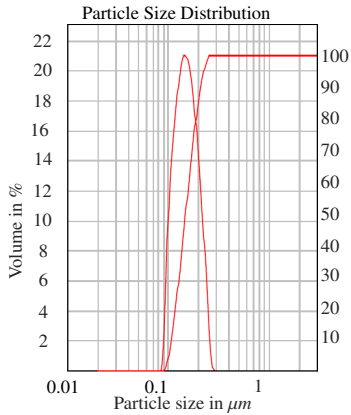}
	\captionof{figure}{Particle size distribution of the used PCS.}
	\vspace{-6pt}
	\label{fig:part_size_distr}
\end{figure}

The used PCS is a paraffin in water dispersion with a mass fraction of approximately $30\%$ paraffin, i.e., $w_\mathrm{P} = 0.3$. A mixture of $50\%$ Eicosan and $50\%$ Docosan is used as paraffin blend to produce the PCS. The test-bed requires $700~\unit{l}$ including the volume of both thermal storages and the piping system. 
The used PCS was produced in eight batches. Figure \ref{fig:dsc_meas} shows the DSC measurement of batch 3 as an example. 
The DSC measurement indicates a latent heat of $\Delta h_f = 49~\unit{kJ/kg}$. The freezing as well as the melting peak occurs in the temperature range between 24 and 42 $^\circ C$. Methods to reduce hysteresis were not applied.
Figure \ref{fig:part_size_distr} provides the information about the paraffin particle size distribution in the PCS. The PCS is stabilized with surfactants to suppress phase separation. The particle size ranges from 0.09 to 0.3 $\unit{\mu m}$.

\begin{figure}[ht]
	\small
	\def\svgwidth{0.45\textwidth} 
	\scriptsize{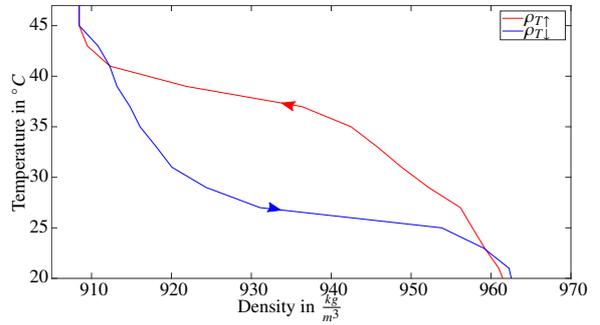}
	\captionof{figure}{Measured density of the used PCS. The red and blue line show the density for rising or falling temperature, respectively.}
	\vspace{-6pt}
	\label{fig:rho_pcs_map}
\end{figure}

The density map of the PCS, which is required to calculate stored thermal energy according to~\eqref{eq:latent_heat}, is shown in Figure \ref{fig:rho_pcs_map}. 

The bottom part of Table \ref{tab:plant_dimensions} refers to the constraints on power and heat flows required in~\eqref{eq:OCP:con_heatgen}-\eqref{eq:OCP:con_elecpower}. The values for the heating rod, grid and battery power follow directly from the manufacturers data sheets. 
The heat pump can be either switched off or operated in the range from $1~\unit{kW_\mathrm{el}}$ to $3.7~\unit{kW_\mathrm{el}}$.
The maximum heat generated by the heat pump is linked to the listed power consumption by~\eqref{eq:P_HP}, where the coefficient of performance depends on the ambient temperature and is calculated according to the heat pump map supplied by the manufacturer \cite{ste_wpl15as}. 

\subsection{Real-time implementation of predictive control}
\label{subsec:RT_impl}
The energy management including the supervisory MPC algorithm runs on a custom embedded hardware with an ARM Cortex-A8 processor with $1~\unit{GHz}$, $512~\unit{MB}$ RAM and a NEON floating-point accelerator. The device is equipped with a module that is connected to the components of the electrical subsystem such as solar charge controller and bidirectional inverter via bus communication.
A proprietary heat pump management system provides the low level control of the hydraulic subsystem (see Figures \ref{fig:heat_gen_schematic} and \ref{fig:th_load_schematic}) \cite{ste_wpm}.
Set-points determined by the predictive controller are communicated via a Modbus TCP connection between the energy management hardware and heat pump management system. This interface allowed for setting heat pump mode and thermal power according to the formal system description in~\eqref{eq:OvSysModel}, i.e., it allows specifying $q_\mathrm{HP,SH}$ and $q_\mathrm{HP,DHW}$.
The three discrete stages of the heating rod are activated by relays. 
The inverter power $P_{\rm G}^\mathrm{dem,sup}$ serves as set-point for electrical power. The optimal setting of  $P_{\rm G}^\mathrm{dem,sup}$ also adjusts battery power $P_{\rm B}^\mathrm{ch,dis}$ optimally, due to constraint~\eqref{eq:ElecCoupling}.
Similarly, the optimal value for $q_{\rm SH}$ does not need to be applied explicitly, as it will appear automatically by hydraulic correlations.

\begin{figure}[ht]
\centering
\def\svgwidth{0.46\textwidth} 
\scriptsize{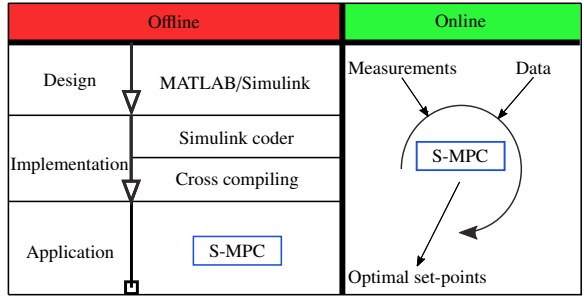}  
\caption{Toolchain for offline design and implementation and scheme of online controller application.}
\label{fig:toolchain}
\end{figure}

Figure \ref{fig:toolchain} summarizes the offline and online steps performed to implement and run the supervisory MPC algorithm.
The controller was designed and implemented in MATLAB/Simulink. C-Code was generated with the Simulink Coder and compiled to the supervisory MPC application.
The application runs periodically on the embedded control platform using a customized Debian Linux as operating system. 
\begin{table}[ht]
\centering
\caption{Assignment of model variables in~\eqref{eq:OvSysModel} to measured real-world quantities, compare Figure \ref{fig:SysStruct} and Figure \ref{fig:electrical_network} and equations~\eqref{eq:E_SH},~\eqref{eq:E_DHW},~\eqref{eq:BatterySoC},~\eqref{eq:q_L_SH}.} 
\vspace{-4pt}
\small
\begin{tabular}{|C{1.15cm}C{1.85cm}C{3.4cm}|}
\hline 
Model variable & Measurement & Description \\
\hline
\multirow{4}{*}{$E_{\rm SH}$} & $T_{\rm center}$ & Tank center temperature \\ 
 & $T_{\rm bottom}$ & Tank bottom temperature \\ 
 & $p_{\rm center}$ & Tank center pressure \\
 & $p_{\rm bottom}$ & Tank bottom pressure \\ \hdashline
\multirow{2}{*}{$E_{\rm DHW}$} & $T_{\rm top}$ & Tank top temperature\\ 
  & $T_{\rm center}$ & Tank center temperature \\ \hdashline
\multirow{1}{*}{$E_{\rm B}$} & $\text{SoC}_{\rm B}$ &  Battery state of charge\\ 
 \hline
 \hline
 \multirow{3}{*}{$q_{\rm L,SH}$}& $\dot{V}_{\rm L,SH}$ & SH sink volume flow \\
 & $T_{\rm L,SH}^{\rm dem}$ & Temperature of SH withdrawal\\
 & $T_{\rm L,SH}^{\rm sup}$ & Cold water supply temperature\\ \hdashline
  \multirow{3}{*}{$q_{\rm L,DHW}$}& $\dot{V}_{\rm L,DHW}$ & DHW sink volume flow \\
 & $T_{\rm L,DHW}^{\rm dem}$ & Temperature of DHW withdrawal \\
 & $T_{\rm L,DHW}^{\rm sup}$ & Cold drinking water supply temperature\\
 \hline
\end{tabular}
\label{tab:meas_calc_variables}
\vspace{-6pt}
\end{table}
The online steps can be summarized as follows. First, measurements and data files are collected. The measurements are used to calculate current states $\mathbf{x}(k)$ and disturbances $\mathbf{d}(k)$ (cf. Table \ref{tab:meas_calc_variables}). The data files include the prediction data described in Sec. \ref{subsec:weath_load_forec} and supervisory MPC parameters, for example tuning parameters and set points in~\eqref{eq:OCP:obj}. 
We can easily tune the supervisory MPC to accomodate changing objectives by manipulating these tuning parameters during run-time.
Gathering all available information,~\eqref{eq:QP} is updated and solved. As a solver, we implemented the open source IPOPT code \cite{waechter2006} and parametrized it to speed up the solution of quadratic programs. After the post processing step sketched in Section \ref{subsec:implementational_aspects}, we gain a suitable, realizable representation of the optimal schedule along the control horizon, of which we send the first element as set-point to the plant.

\section{Experimental results}
\label{sec:ExpRes}

\subsection{Controller setup}
\label{subsec:SetupMPC}
The sampling time for the supervisory MPC was $15~\unit{min}$. 
The results show that this sampling time is sufficient to achieve the control objective, since the control performance depends on the mid to long term behavior of the plant. 
Measurements were acquired every two minutes for monitoring purposes. This resolution is suitable for capturing the states of charge of the storage devices.
The electric power flows can fluctuate at higher frequencies. Hence, we filter these values for the presentation.

We chose to maximize the PV self-consumption as control objective for the experiments. This implies a minimization of the PV power fed into the grid and requires to plan thermal and electric storage dispatch in advance. 
To this end, the reference values in \eqref{eq:CostFunction_Outputs} are set to $E_{\rm SH}^{\rm set} = E_{\rm SH}^{\rm max}=\mathrm{const.}$, $E_{\rm DHW}^{\rm set} = E_{\rm DHW}^{\rm max}=\mathrm{const.}$ and $E_{\rm B}^{\rm set} = E_{\rm B}^{\rm max} =\mathrm{const.}$. 
The tuning parameters are chosen according to Table \ref{tab:smpc_tuning}, where $r_{\rm dem}(t)=\mathrm{const.}$ and $r_{\rm sup}(t)=\mathrm{const.}$ denote actual electricity cost and compensation, respectively.
\begin{table}[h!]
\center 
\caption{Cost function parameters for maximum PV self-consumption.}
\small
\begin{tabular}{|l|C{2.4cm}|C{2.4cm}|}
\hline
$r_{\star}(k)$ & $t\in \left[6, 22\right] \mathrm{h}$ & $t\in \left[22, 6\right] \mathrm{h}$ \\
\hline
$\rm E,SH$ & 3 & 0.01 \\ 
$\rm E,DHW$ & 5 & 0.5 \\ 
$\rm E,BLD$ & 1 & 0.1 \\ 
$\rm E,B$ & 3 & 1 \\
\hline
$\rm HP,SH$ & \multicolumn{2}{c|}{\mbox{$^5$/$_{COP(t)}$}} \\ 
$\rm HP,DHW$ & \multicolumn{2}{c|}{\mbox{$^{20}$/$_{COP(t)}$}} \\ 
$\rm HR$ & \multicolumn{2}{c|}{$250$} \\ 
$\rm B$ & \multicolumn{2}{c|}{$\begin{cases} 1\qquad,~\text{charge} \\ 1\qquad,~\text{discharge} \end{cases}$} \\ 
$\rm G$ & \multicolumn{2}{c|}{$\begin{cases} \begin{cases} 1e^3\cdot r_{\rm dem}(t),~P_{\rm PV} \leq 1\unit{kW}\\1e^4\cdot r_{\rm dem}(t),~P_{\rm PV} > 1\unit{kW}\end{cases},~ \rm demand \\ 10\cdot r_{\rm sup}(t)\qquad~~,~\rm supply \end{cases}$} \\
\hline
\end{tabular}
\vspace{-10pt}
\label{tab:smpc_tuning}
\end{table}

\subsection{Experimental validation}
\label{sec:Perf_SMPC}
\begin{figure}[ht]
\centering
\def\svgwidth{0.46\textwidth} 
\scriptsize{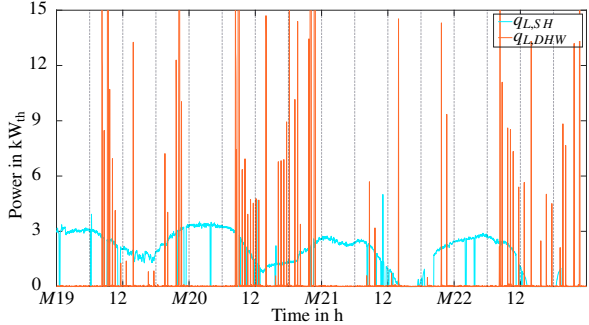}
\caption{Thermal load demand for March $19th$ to March $22nd$. Blue and red line show space heating and domestic hot water demand, respectively.}
\label{fig:therm_load_demand}
\vspace{-5pt}
\end{figure}

We investigate the supervisory MPC performance for a four day time-span during spring season with central German climate conditions.
The mean ambient temperature and mean daily PV generation amount to $7.2~^\circ C$ and $19.1~\unit{kWh}$, respectively. 
The thermal load demand is shown in Figure \ref{fig:therm_load_demand}. The DHW load matches load profiles from EU commission regulation \cite{eu_ver814_2013}. 
\begin{figure*}[ht]
\centering
\begin{subfigure}[c]{1\textwidth}
\centering
\def\svgwidth{1\textwidth} 
\scriptsize{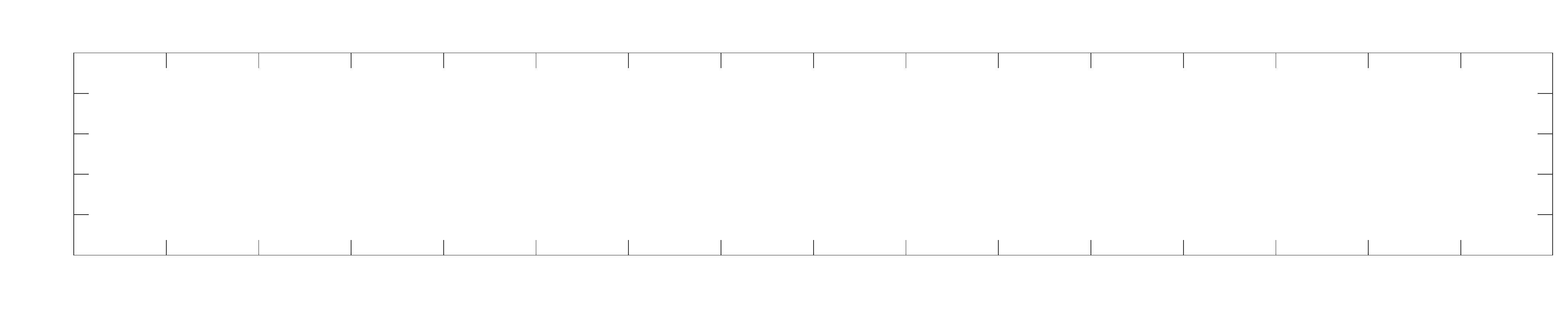}
\subcaption{Temperatures measured in the storage tank. The red line depicts the temperature at the top, the blue line in the middle and the yellow line at the bottom, respectively. The dashed lines mark the bounds of the phase change range.}
\label{fig:storage_res}
\end{subfigure}
\begin{subfigure}[c]{1\textwidth}
\centering
\def\svgwidth{1\textwidth} 
\scriptsize{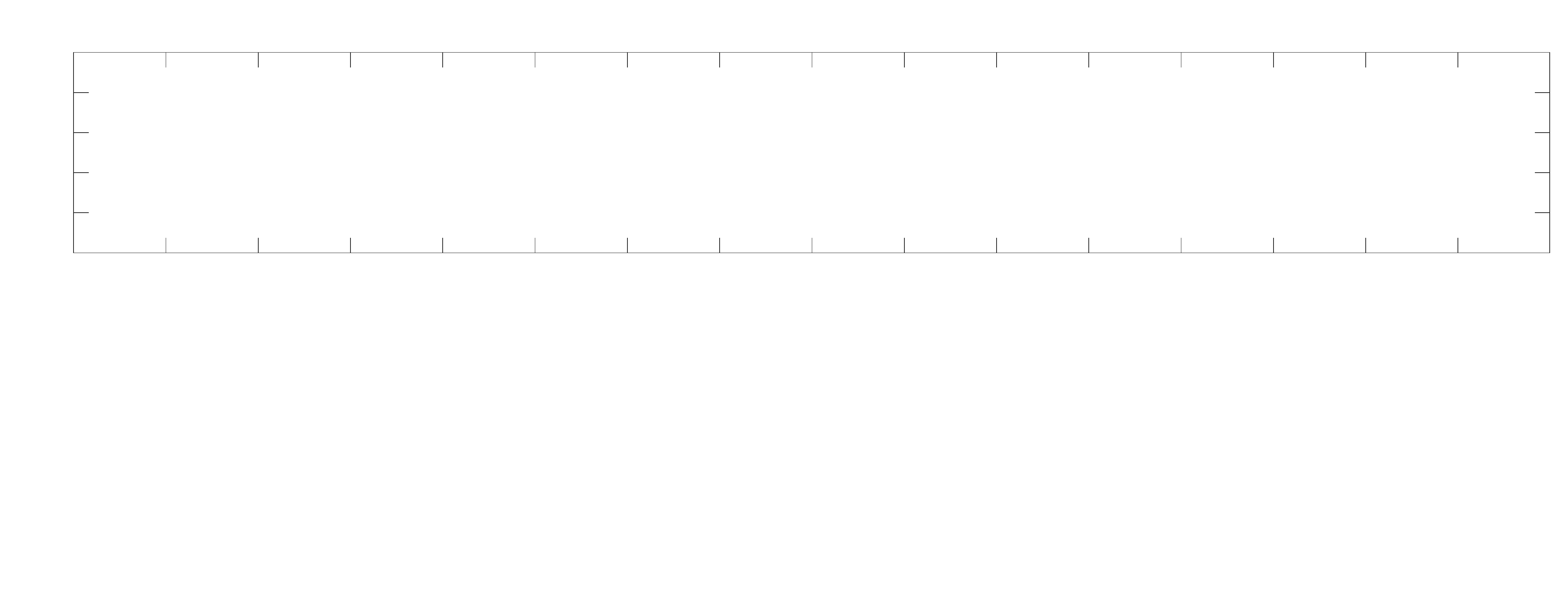}
\subcaption{Details on heat pump operating behavior. The top figure shows supply and return temperature, the bottom figure shows measured power demand and heat generation calculated by~\eqref{eq:q_HP}.}
\label{fig:heat_pump_res}
\end{subfigure}
\vspace{-5pt}
\caption{Experimental results for supervisory MPC of the hydraulic part of the plant, March 2019.}
\vspace{-5pt}
\label{fig:exp_res_thermal}
\end{figure*}
The SH heat load is more evenly distributed showing maximum SH load during early morning hours and minimal load during afternoon hours. The SH load results from the heat demand necessary to maintain a constant room temperature of $21~^\circ C$. 

Figure \ref{fig:storage_res} shows that the  supervisory MPC is able to operate the lower part of the storage within the phase change range marked by the dashed lines most of the time. This enforces frequent melting and solidification of the PCS and thus good exploitation of its latent heat capacity. 
The upper part of the storage is always held above $50~^\circ C$. This requirement is enforced by constraint~\eqref{eq:OCP:StorageBounds} to guarantee a suitable temperature for DHW supply. 

As the storage is filled with a PCS, it would be expected to observe temperature plateaus in Figure \ref{fig:storage_res}.
To explain why these plateaus do not occur, we stress again that the individual particles of the PCS are not located permanently in the storage but are pumped. Consequently, the phase change occurs also within the piping and not only at a constant geometric location inside the storage tank.
It will be shown in Section \ref{subsec:HPandPCS_results} that the latent capacity of the PCS is indeed used effectively.

The operating behavior of the heat pump is evident from Figure \ref{fig:heat_pump_res}. 
Supply temperatures larger than $40~^\circ C$ correspond to the heat pump providing heat for DHW. Lower temperatures indicate the heat pump provides heat for SH. 
The lower plot indicates the relation between power consumption and heat generation.

\begin{figure*}[ht]
\centering
\def\svgwidth{1\textwidth} 
\scriptsize{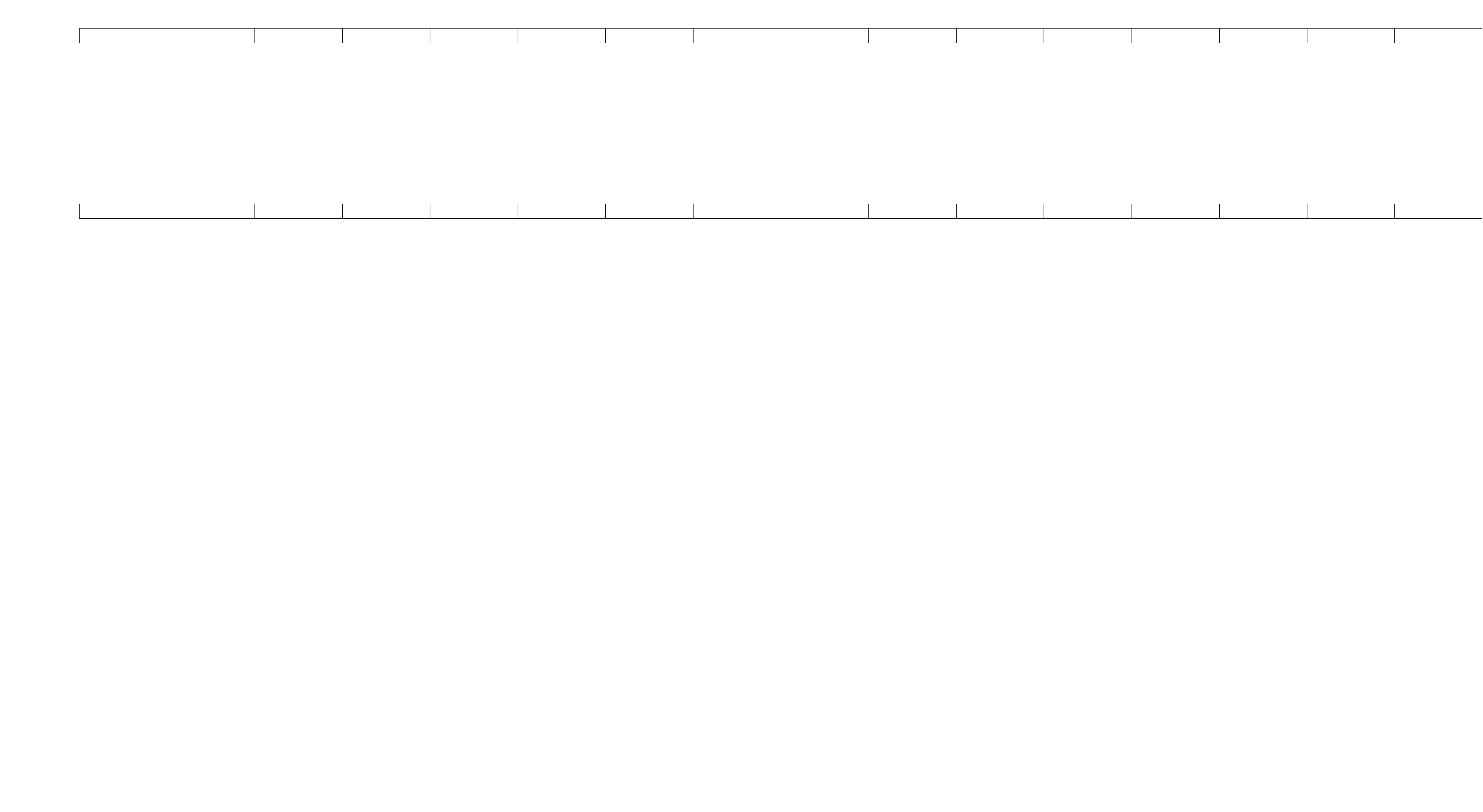}
\vspace{-5pt}
\vspace{-5pt}
\caption{Details on power flows and battery usage, March 2019. }
\vspace{-13pt}
\label{fig:elec_res}
\end{figure*}
Figure \ref{fig:elec_res} shows the electric power flows and the state of charge of the battery. The first two days show a low PV power output. The state of charge is kept at a low value above the minimum feasible level of $35 \%$. 
There was a higher PV generation at the last two days. Consequently, the battery is being operated at higher levels. By anticipatory dispatch of the battery storage, combined with the simultaneous heat generation from PV power, the controller almost completely prevented feeding electric energy into the grid, which can be mainly seen at noon. Heat generation and power storage occur in parallel, such that the battery storage is not charged completely while the sun is shining, i.e., until around 6 p.m. The PV-self consumption was $91.8 \%$.

Additionally, grid demand was kept at a low level, while some peaks exist where the heat pump ran in DHW mode. Considering the second day in Figures \ref{fig:heat_pump_res} and \ref{fig:elec_res}, a relatively high DHW demand coincides with a moderate PV power generation of the previous day. This causes grid demand as a consequence of the low battery state of charge. 

\subsection{Heat load shifting and grid support}
 \label{sec:loadshift_grid_results}
\begin{figure}[ht]
\centering
\def\svgwidth{0.47\textwidth} 
\scriptsize{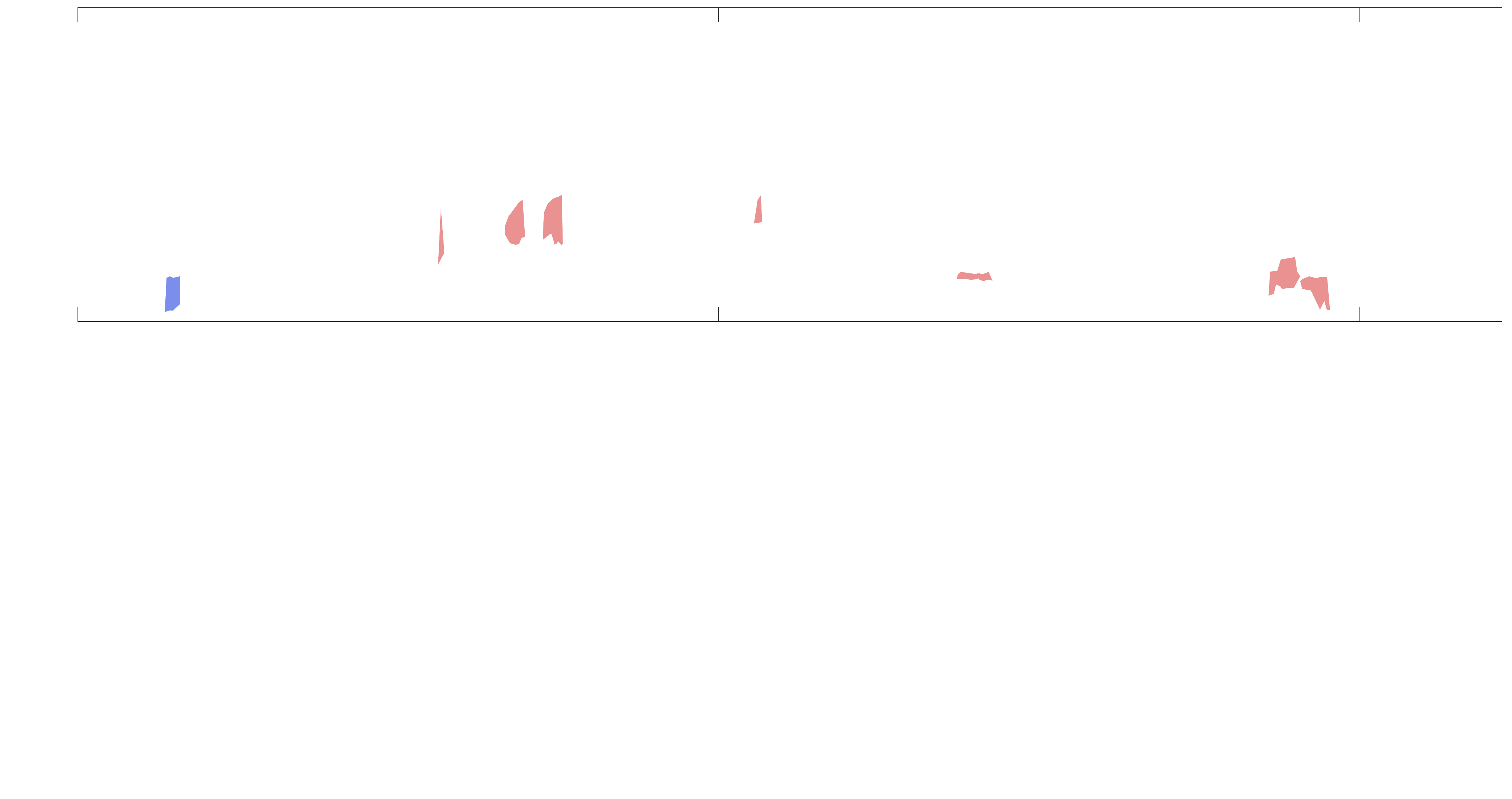}
\vspace{-5pt}
\captionof{figure}{Grid relief during PV generation period achieved by the supervisory MPC, illustrated for March 20st, 2019. The blue and red shaded areas in the upper subplot mark the additional power drawn from battery and grid, respectively.}
\vspace{-5pt}
\label{fig:illu_grid_rel}
\end{figure}

Operation strategies for maximum PV self-consumption often lack grid friendly performance and large feed-in peaks are common. While simple management strategies already are able to reduce these peaks \cite{moshoevel2015}, 
our system achieves further improvements. 
Thermal load shifting can be observed in Figures \ref{fig:exp_res_thermal} and \ref{fig:elec_res}. In times of PV generation,  $52.8 \%$ of the overall heat has been produced, while only $37.3 \%$ of heat load occurred then.
Especially on the latter two days, it is evident that the controller stores electrical and thermal energy from PV generated power. Consequently, in the experiment, $45.19~\unit{kWh}$ or $15.5 \%$ of heat generation has been shifted to times with PV generation. 
The direct use of PV generated power was $37 \%$, which accounted for $74 \%$ of heat generation during this time. 
The remaining power was drawn from the battery or the grid. More precisely, $10.29~\unit{kWh}$ or about $34 \%$ of the $29~\unit{kWh}$ of total grid demand have occurred in sunny times. If this number is scaled to regions with multiple domestic PV installations, such a system operation provides significant grid relief in times of high solar penetration, see Figure \ref{fig:illu_grid_rel}.
In addition to the power withdrawal from the grid, we also observe that the battery was discharged at sunny times. 
Consequently, the controller extends the time span in which the battery storage can be charged from PV power.

In summary the supervisory MPC operates both storage domains within useful intervals. Thus, it increases the amount of self-generated power which can be used on-site.
This behavior results from the tuning parameters (see Table \ref{tab:smpc_tuning}). 
Decreasing the penalties on grid action or linking it to a time-varying price profile allows for grid supply during PV generation if there is a suitable incentive.

\subsection{Heat pump operation and phase change slurry}
\label{subsec:HPandPCS_results}
The heat pump efficiency and the increased heat capacity of the storage
are further key performance indicators of the proposed setup. 
The supervisory MPC usually operates the heat pump at long run-times of around $1~\unit{h}$, see Table \ref{tab:COPvsRuntime}. This is beneficial because the heat pump efficiency increases with longer run-times. 
\begin{table}[ht]
\center
\caption{Heat pump COP in SH operation mode related to run-time per activation. Run-times smaller than 0.5 hours were not planned by supervisory MPC operation but occur due to heat pump specific behavior like defrosting, which cannot be scheduled explicitly.}
\begin{tabular}{|c|c|c|}
\hline
Run-time & $COP_{\mathrm{SH}}$ & Share of time \\
\hline
$ 0 - 0.5 h$ & $3.5$ & $12 \%$ \\
$ 0.5 - 0.75  h$ & $3.2$ & $12 \%$ \\
$ 0.75 - 1 h$ & $4.1$ & $39 \%$ \\ 
$ 1 - 1.25 h$ & $4.2$ & $27 \%$\\
$ 1.25 - 1.5 h$ & $3.9$ & $9 \%$\\
\hline
\end{tabular}
\vspace{-5pt}
\label{tab:COPvsRuntime}
\end{table}

Table \ref{tab:COPvsTamb_temp} shows that the efficiency is higher at lower ambient temperatures, which is counter-intuitive, as the coefficient of performance usually increases with increasing temperature.
It is evident, however, that longer run-times are required at lower ambient temperatures, as higher thermal loads occur in these cases. 
Hence, the disadvantage of running the heat pump with low air-source temperature is compensated by the long run-times.
\begin{table}[ht]
\center
\caption{Heat pump efficiency in SH operation mode related to ambient temperature.}
\begin{tabular}{|c|c|c|}
\hline
$T_{amb}$ & $COP_{\mathrm{SH}}$ & Share of time \\
\hline
$ \leq 0~^\circ C$ & $4.0$ & $21 \%$ \\
$ 0-5  ^\circ C$ & $4.4$ & $36 \%$ \\
$ 5-10  ^\circ C$ & $4.4$ & $25 \%$ \\
$ 10-15  ^\circ C$ & $3.8$ & $12 \%$  \\
$ 15-20  ^\circ C$ & $2.6$ & $4 \%$  \\
$ \geq 20  ^\circ C$ & $2.9$ & $1 \%$ \\
\hline
\end{tabular}
\vspace{-5pt}
\label{tab:COPvsTamb_temp}
\end{table}

\begin{figure*}[ht]
\centering
\def\svgwidth{0.9\textwidth} 
\scriptsize{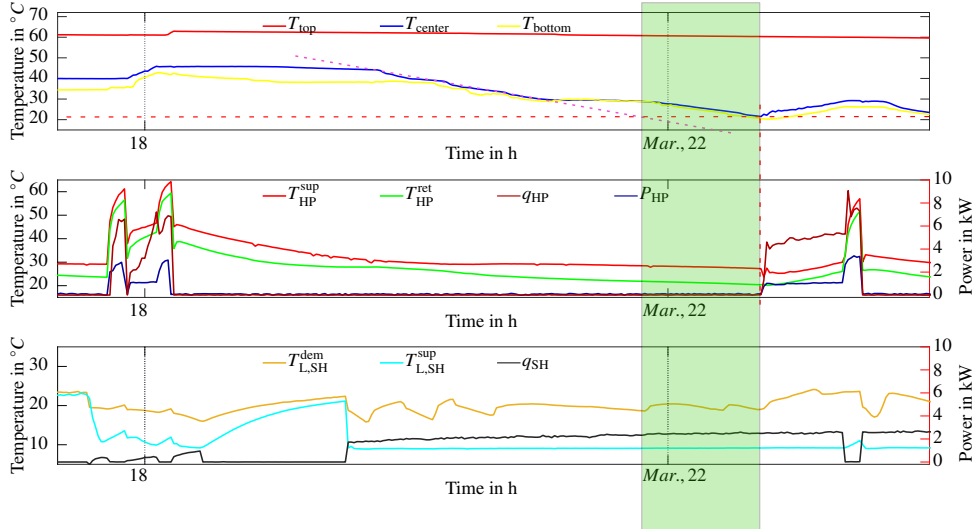}
\vspace{-5pt} 
\caption{Indication of the heat load shifting effect in the storage induced by the phase change slurry. The sloped dashed line in the upper subplot projects the sensible heat discharge for conventional water as storage medium. The green shaded area indicates the time span by which the heat pump activation was delayed. The lower two subplots show heat generation and heat demand, respectively. }
\label{fig:PCS_slow_temp_decrease}
\end{figure*}

The PCS does not only increase the heat capacity of the storage tank, but it also has a positive effect on the overall system operation.
In general, the charge or discharge of latent heat causes the storage temperature to be almost constant for longer periods during the phase change.
This is evident from the time around midnight of March $22nd$ in Figure \ref{fig:PCS_slow_temp_decrease}. The temperature in the storage decreases linearly, until it reaches about $29.6~^\circ C$, which equals the upper bound of the solidification temperature range (cf. Figure \ref{fig:dsc_meas}). At this point, it remains almost constant for more than one hour, before the temperature decrease continues. The bottom plot shows that the SH load is almost constant, i.e., heat is removed from the storage without adding heat by the heat pump (center subplot). Precisely, $2.3~\unit{kWh}$ load demand was fulfilled without significant temperature decrease in the storage, which delayed the next heat pump activation by $1.5~\unit{h}$ compared to the operation with water. Hence, the phase change slurry yields further flexibility potential. 
The specific properties of the phase change slurry can also be observed in periods of charging the thermal storage. This is evident from the data for March, $20th$, $2~\unit{pm}$ to $6~\unit{pm}$. During this time-span, $8.9~\unit{kWh}$ of heat are generated by the heat pump while $4.3~\unit{kWh}$ are demanded by the heating system and $4.6~\unit{kWh}$ are stored in the thermal storage. The total thermal storage capacity is $8.4~\unit{kWh}$.
The temperature shift of $12.1~\unit{K}$ corresponds to $3~\unit{kWh}$ heat that is stored sensibly. Hence, the amount of latent heat stored is approximately $E_{\mathrm{lat}} = 1.6~\unit{kWh}$. This increases the heat capacity that can be used to store PV generated power compared to the operation with water.
The theoretical maximum amount of latent heat in the thermal storage is $3.9~\unit{kWh}$. During the analyzed period, there was a phase transition of approximately $40 \%$ of the paraffin.

%

\section{Conclusion and Outlook}
\label{sec:ConcOut}
We presented a supervisory model predictive control approach for an electrical heating system featuring a phase change slurry as heat transfer and heat storage medium. 
We demonstrated the applicability of our control approach with an experimental test-bed 
and showed heating systems composed of standard components can be operated with a phase change slurry as a heat transfer fluid.
The control system achieves the desired high PV self-consumption, operates the heat pump efficiently, and provides a significant load shifting potential. 
Moreover, it meets the real-time requirements even when implemented on low-cost hardware, and it can be added to a existing low-level controllers as a piggyback supervisory system without affecting any safety-critical function.  

Future work will address cost functions that take dynamic electricity prices into account. Furthermore, aging models for the heat pump and for the electric battery will be included to ensure an optimal operation over the entire system lifecycle.


\vspace{-5pt}

\appendix
\section{Heat pump heat generation with phase change slurry} \label{sec:app:heat_gen_and_release_PCS}
\begin{figure}[ht]
\centering
\def\svgwidth{0.47\textwidth}
\small{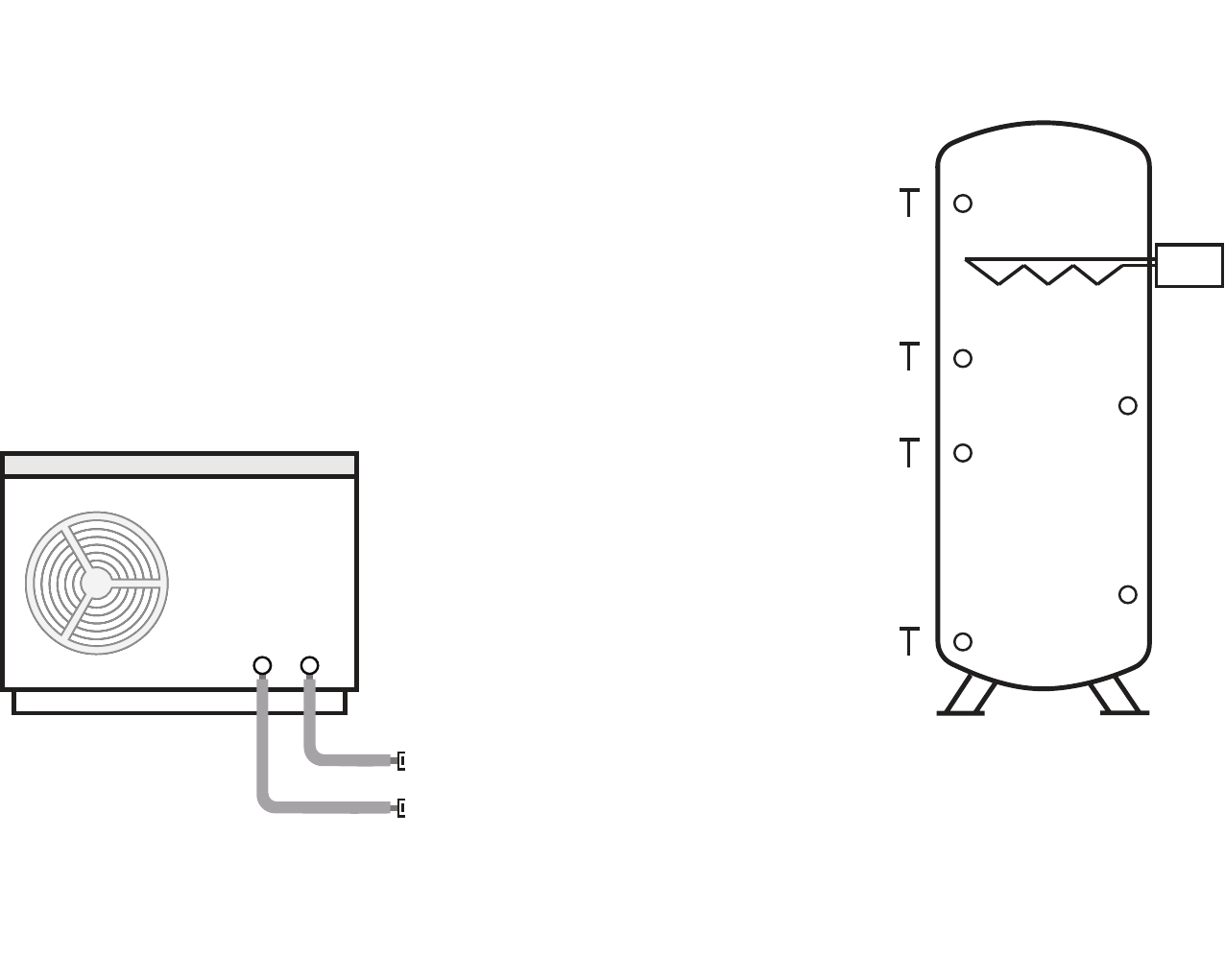}
\vspace{-10pt}
\caption{Heat generation circuit.}
\vspace{-5pt}
\label{fig:heat_gen_schematic}
\end{figure}

The hydraulic circuit for heat generation is shown in Figure \ref{fig:heat_gen_schematic}.
Since the heat transfer fluid is a PCS, the heat pump simultaneously heats up the fluid and melts the paraffin particles.
Equation~\eqref{eq:latent_heat} introduced in Section~\ref{subsec:PCS} cannot be used to determine the fraction of latent heat, since no pressure measurements are available in the piping.
We approximate the generated heat by replacing the calculated specific heat capacity $c_{p,\rm PCS}$ by a nonlinear map $c_{ p,\rm PCS}^\mathrm{approx.}\left(T\right)$ \cite{mehling2008}.
The map was derived by grey box estimation techniques and accounts for the measurement peaks in Figure \ref{fig:dsc_meas}.
The generated heat can then be calculated from
\small
\begin{align}
\label{eq:q_HP}
q_\mathrm{HP,SH} &= \rho_{\rm PCS}\left(T_\mathrm{HP,SH}^{\rm sup}\right) \cdot \dot{V}_\mathrm{HP,SH} \cdot c_{p,\rm PCS}^\mathrm{approx.}\left(T_\mathrm{HP,SH}^{\rm sup}\right) \cdot \Delta T_\mathrm{SH}, \\
q_\mathrm{HP,DHW} &= \rho_{\rm PCS,liq} \cdot \dot{V}_\mathrm{HP,DHW} \cdot c_{p,\rm PCS} \cdot \Delta T_\mathrm{DHW},
\end{align}
\normalsize
where $\Delta T_\mathrm{SH} = \left( T_\mathrm{HP,SH}^{\rm sup} - T_\mathrm{HP,SH}^{\rm ret} \right)$, $\Delta T_\mathrm{DHW} = \left( T_\mathrm{HP,DHW}^{\rm sup} - T_\mathrm{HP,DHW}^{\rm ret} \right)$ and where the density $\rho_{\rm PCS}\left(T\right)$ depends on the temperature of the fluid (see Figure \ref{fig:rho_pcs_map}).
The current operating mode of the heat pump determines if $q_\mathrm{HP}=q_\mathrm{HP,SH}$ or $q_\mathrm{HP}=q_\mathrm{HP,DHW}$.

\section{Parameter identification} \label{sec:app:parameter_identification}
The scalar parameters $\alpha_i, i=1,\dots,4$ represent storage losses, $\nu$ measures the thermal heat flow between upper and lower part of the storage and $\beta_i,i=1,\dots,8$ collect the charging and discharging efficiencies.
We calculate the thermal loss parameters $\alpha_i,~i=1,2,3$ based on total storage capacities and nominal conditions. All other parameters of the linear model are fitted to measurement data with matlab's system identification toolbox. The values listed in Table \ref{tab:model_parameters} are identified based on ten measurements at different ambient conditions. The mean goodness of the fit was $59 \%$ and $84 \%$ for the parameters of thermal and electrical subsystem, respectively. 
\begin{table}[ht]
\centering
\caption{Calculated and identified loss parameters and efficiencies.}
\begin{tabular}{| l | l | | l | l |} \hline \label{tab:model_parameters} 
Loss param. & Value & Eff. param. & Value \\ \hline
 $\alpha_1$ & $0.99949$ &$\beta_1$ & $0.275$  \\
 $\alpha_2$ & $0.9979$ &$\beta_2$ & $0.298$ \\
 $\alpha_3$ &  $1$ &$\beta_3$ & $0.192$ \\
 $\alpha_4$ & $0.9991$ &$\beta_4$ & $0.248$ \\
 $\nu$  & $0.003$ &$\beta_5$ & $0.339$ \\
 & & $\beta_6$ & $0.298$ \\
& & $\beta_7$ & $0.223$ \\
& & $\beta_8$ & $0.205$ \\
 \hline
\end{tabular}
\end{table}

\section{Calculation of stored energy}
\label{sec:app:stored_energy}
\begin{figure}[ht]
\centering
\small
\def\svgwidth{0.25\textwidth} 
\small{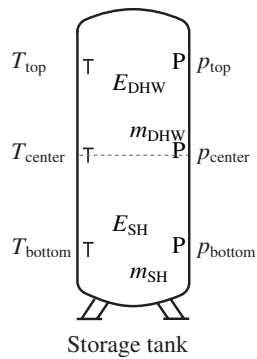}
\caption{Storage tank with measured quantities.}
\label{fig:heat_storage}
\end{figure}

The storage tank is equipped with temperature and pressure sensors as sketched in Fig. \ref{fig:heat_storage}.
As the phase change range was selected below the DHW temperature level, latent heat only contributes in the lower part. Let $T_\mathrm{SH}$ and $T_\mathrm{DHW}$ be the lowest temperature values usable for SH and DHW supply and $m_\mathrm{\left(SH,DHW\right)}$ and $\rho_\mathrm{\left(SH,DHW\right)}$ denote the mass and density of PCS in the lower and upper part of the storage, respectively.
With~\eqref{eq:q_PCS} -~\eqref{eq:latent_heat} and $\rho_\mathrm{SH} = f\left(p_\mathrm{center}, p_\mathrm{bottom}\right)$, the total amount of heat stored can be calculated from
\begin{equation}  \label{eq:E_SH}
E_\mathrm{SH} = m_\mathrm{SH}\cdot c_{p,\mathrm{PCS}} \cdot \left( T_\mathrm{center} - T_\mathrm{SH} \right) + E_\mathrm{lat}(\rho_\mathrm{SH})
\end{equation} 
and
\begin{equation}\label{eq:E_DHW}
E_\mathrm{DHW} =m_\mathrm{DHW} \cdot c_{p,\mathrm{PCS}} \cdot \left( T_\mathrm{top} - T_\mathrm{DHW} \right).
\end{equation}

\section*{Acknowledgment}
Support by the German Federal Ministry for Economic Affairs and Energy under grant 03ET1274A-D is gratefully acknowledged. Thanks go to Christian Kolbe and Christoph Wiebicke from HPS Home Power Solutions for their continuous support and contribution.
\vspace{-5pt}
\bibliographystyle{elsarticle-num} 
\bibliography{Loehr2020Arxiv}



\end{document}